\documentclass[twocolumn,english,prb]{revtex4}
\pdfoutput=1
\usepackage[T1]{fontenc}
\usepackage[latin1]{inputenc}
\usepackage{amsmath}
\usepackage{graphicx}
\usepackage{amssymb}
\makeatletter
\usepackage{float}
\makeatletter
\makeatletter
\makeatother
\makeatother

\usepackage{babel}
\makeatother
\begin{document}
\newcommand{\be}{\mathbf{E}}

\newcommand{\bj}{\mathbf{j}}

\newcommand{\br}{\mathbf{r}}

\newcommand{\zhat}{\hat{\mathbf{z}}}

\newcommand{\sym}{a^{\mathrm{sym}}}

\newcommand{\asym}{a^{\mathrm{asym}}}

\newcommand{\sd}[1]{#1^{\prime\prime}}

\newcommand{\fd}[1]{#1^{\prime\prime\prime\prime}}

\title{Are Microwave Induced Zero Resistance States Necessarily Static?}

\author{Ilya G. Finkler and Bertrand I. Halperin}

\affiliation{Department of Physics, Harvard University, Cambridge, Massachusetts
02138}

\begin{abstract}
We study the effect of inhomogeneities in Hall conductivity on the
nature of the Zero Resistance States seen in the microwave irradiated
two-dimensional electron systems in weak perpendicular magnetic fields,
and we show that time-dependent domain patterns may emerge in some
situations. For an annular Corbino geometry, with an equilibrium charge
density that varies linearly with radius, we find a time-periodic
non-equilibrium solution, which might be detected by a charge sensor,
such as an SET. For a model on a torus, in addition to static domain
patterns seen at high and low values of the equilibrium charge inhomogeneity,
we find that, in the intermediate regime, a variety of nonstationary
states can also exist. We catalog the possibilities we have seen in
our simulations. Within a particular phenomenological model, we show
that linearizing the nonlinear charge continuity equation about a
particularly simple domain wall configuration and analyzing the eigenmodes
allows us to estimate the periods of the solutions to the full nonlinear
equation. 
\end{abstract}
\maketitle

\section{Introduction}

A novel {}``zero-resistance state'' (ZRS) was observed a few years
ago, when a two-dimensional electron gas (2DEG), subjected to a weak
magnetic field, was also irradiated with microwaves.\cite{ZRS-exp1,ZRS-exp2,ZRS-exp3,ZRS-exp4}
In the ZRS state, the dc resistance measured in a Hall bar experiment
appeared to vaniish, while correspondingly, the longitudinal conductivity
$\sigma_{xx}$, measured in a Corbino geometry appeared to be zero.
A related phenomenon of microwave-induced resistance oscillations
had been observed earlier in samples of lower quality, in which there
are changes in the dc resistance produced by the microwave irradiation,
which depend in an oscillatory manner on the ratio of the microwave
frequency to the cyclotron frequency of the 2DEG, but where the resistance
was not driven down to zero.\cite{ZRS-pre1,ZRS-pre2}

Following the discovery of the ZRS, a phenomenological explanation
was put forward.\cite{ZRS-macro} It was assumed that for microwave
power above a certain threshold, for an appropriate sample, in certain
ranges of the magnetic field and microwave frequency, the microwave-induced
resistance oscillations would increase to the point where the differential
longitudinal conductivity would become negative at zero dc electric
field, and it was noted that in this case that the zero-electric-field
solution would necessarily be unstable. It was argued that the sample
would spontaneously break up into a pattern of domains, separated
by domain walls. Within each domain, the magnitude of the electric
field would be a constant, $E_{c}$, dependent on the strength of
the microwave field and other parameters, which satisfies a condition
that the longitudinal current generated by the field is zero. The
direction of the electric field would change discontinuously at a
domain wall, and might also vary continuously within a domain. It
was assumed that the domain walls could move rather easily, so that,
\textit{e.g.}, if a finite voltage is applied between the inner and
outer edges of a Corbino sample, the system would respond by a displacement
of domain walls, without causing the magnitude of the field to deviate
from the critical value $E_{c}$, and without inducing a longitudinal
current in the sample.

If one accepts this general picture, an obvious question is what determines
the domain pattern in any given sample? In fact, one may question
whether there is necessarily a static domain pattern at all. Since
the system is driven out of equilibrium by the constant absorption
of microwave radiation, it is possible in principle that the system
will enter a time-dependent state, where domain walls may move about
in a chaotic, or possibly quasiperiodic, manner. It is our purpose
here to address this question. We shall argue that time-dependent
states are likely to occur in at least some situations, and we shall
propose one experimental geometry where it should be possible to observe
this time-dependence. The dc properties of a time-dependent state
should be similar to those of a state with static domain walls; in
particular the observed value of $\sigma_{xx}$ should be zero, or
at least very small compared to the conductivity in the absence of
microwave radiation.

At least two different microscopic mechanisms have been proposed to
explain the microwave-induced resistance oscillations at lower microwave
intensities, and to produce the negative zero-field conductance assumed
in the macroscopic model of ZRS: a {}``displacement mechanism''\cite{Assa,Durst,Ryzhii1,Ryzhii2,Ryzhii3,Shi,ZRS-DP3,ZRS-DP4,ZRS-DP5},
in which the absorption of a microwave photon by an electron leads
(for appropriate values of parameters) to a disorder-assisted displacement
of the cyclotron orbit in an up-hill direction with regard to the
local dc electric field; and a {}``population mechanism''\cite{ZRS-DF1,ZRS-DF2},
in which absorption of the microwave photons leads to a population
inversion in the partially filled Landau levels close to the Fermi
level. Although these mechanisms may lead to quite different estimates
for such parameters as the threshold microwave power or for the critical
field $E_{c}$ at a given level of microwave irradiation, it appears
that they give rise to qualitatively similar phenomenological models.
In any case, we shall not make any assumptions about the particular
microscopic mechanism in this work.

In formulating the equations of phenomenological model, it is convenient
to divide the local electric current into a longitudinal or {}``dissipative''
part, and a Hall current which is always perpendicular to the local
field $\be$. In previous work, \cite{ZRS-ours1,ZRS-ours2} we considered
a model where the functional form of the dissipative current could
vary from one place to another in the sample, due to small inhomogeneities
in the doping density, or due to other sources of disorder. However,
we assumed the Hall current to be governed by a linear (in field $\be$)
Hall conductivity, which we took to be uniform throughout the sample.
In this case, we were able to derive a Lyapunov functional, whose
value can only decrease in time, for a system with specified electrochemical
potential at the boundaries. Then the system must eventually reach
a static state, whose potential configuration is at least a local
minimum of the Lyapunov functional, so that oscillatory or chaotic
time-dependent solutions are not possible as a steady state.

If we allow the Hall conductivity to vary from one place to another,
however, it is generally not possible to find a Lyapunov functional
for the system. In this case, there is no minimization principle to
determine the long-time behavior, and there is no guarantee that a
stable time-independent solution exists. In fact, we shall explore
explicitly some simple geometries where, for certain ranges of parameters,
the long-time behavior is time-dependent.

The rest of the paper is organized as follows. In Sec. \ref{sec:Model},
we discuss the phenomenological relation between the dissipative current
and the local field, provide a brief overview of the Lyapunov functional
formalism, then use the local capacitance model to arrive at a continuity
equation that can be solved for the position-dependent electrochemical
potential. We consider the solutions to the continuity equation for
two geometries, the first being Corbino-like, with the potential held
fixed at two edges along one direction and periodic along the second
direction. In a second geometry (torus), periodic conditions along
both directions are used. Having established in Sec. \ref{sec:Model}
that only variations in the equilibrium charge density contribute
to the divergence of the Hall current, in Sec. \ref{sec:Uniform-Hall-Conductance}
we describe, for both geometries, the solutions to continuity equation
for uniform equilibrium charge density. Sec. \ref{sec:Exact-Solution-in}
contains the central result of the paper; we show that, in a Corbino
setup, linearly varying equilibrium charge density would give rise
to periodic solutions. In Sec. \ref{sec:Torus Numerics}, we present
the numerical evidence for the presence of nonstationary solutions
on the torus. The following section, Sec. \ref{sec:Torus-Analytics},
is dedicated to gaining analytical understanding of how solutions
of previous section arise by analyzing the linearized continuity equation.
In Sec. \ref{sec:Simulation-Details}, we provide some details on
how simulations have been carried out and compare the predictions
of the linearized analysis of the previous section with the numerical
simulations. We comment on experimental prospects in our concluding
section, Sec. \ref{sec:Experimental-Prospects}. A discussion of some
alternative assumptions for the Lyapunov function, and their consequences
for the relative costs of different domain-wall configurations, is
given in an Appendix.

\section{\label{sec:Model}Model}

In the presence of an external microwave field, we assume the following
relation between the dc current $\mathbf{j}(\mathbf{r})$ and the
field $\mathbf{E}(\mathbf{r})$:\begin{equation}
\bj=\mathbf{\bj}^{d}(\mathbf{\be},\mathbf{\br})+\sigma_{H}\,\zhat\times\be-\lambda\nabla^{2}\be,\label{eq:curr}\end{equation}
 where $\be\equiv-\nabla V(\br)\equiv\frac{\nabla\mu(\br)}{e}$ and
$\zhat$ is a unit vector normal to the plane. Here, $\mu(\br)$ is
the electrochemical potential and $-e$ is the electron charge. For
the sake of convenience, we will from here on refer to $\be$ as the
electric field and $V(\br)$ as the electrochemical potential. In
Eq. \ref{eq:curr}, we will explicitly allow the Hall conductivity
$\sigma_{H}$ to be position-dependent. The dissipative current $\bj^{d}$
satisfies the condition that the differential dissipative conductivity,
$\sigma_{\alpha\beta}^{d}(\be)\equiv\partial j_{\alpha}^{d}/\partial E_{\beta}$,
is symmetric:\begin{equation}
\sigma_{\alpha\beta}^{d}(\be,\br)=\sigma_{\beta\alpha}^{d}(\be,\br)\label{eq:jdiss}\end{equation}
 The nonlocal (third) term in Eq. (\ref{eq:curr}) implements an ultraviolet
cutoff, which will lead to a finite domain wall thickness, proportional
to the square-root of the parameter $\lambda$, which will be taken
to be small compared to the size of typical domain (indeed, this is
almost a matter of definition of domains phase). In practice, we expect
that the domain wall thickness will be of the order of the cyclotron
radius $l_{c}$, which is of order 1 $\mu\mathrm{m}$ in typical samples
where ZRS is observed.

The vector function $\bj^{d}$ may also depend explicitly on the position
$\br$, due to inhomogeneities in the 2DEG, and its direction may
not be perfectly aligned with $\be$. In previous work, we considered
explicitly the effects of inhomogeneities in $\bj^{d}$ arising from
gradients in the equilibrium electrostatic potential $\phi_{d}$ due
to disorder. In the present paper, however, we shall ignore this complication,
and shall assume that the function $\bj^{d}$ has no explicit dependence
on $\br$, except for a brief discussion at the end.

Eq. (\ref{eq:curr}) will be supplemented by the continuity equation,
\begin{equation}
\dot{\rho}=-\nabla\cdot\bj\,,\end{equation}
 where $\rho$ is the charge density. Writing $\be\equiv-\nabla V(\br)$,
we may relate changes in the electrochemical potential $V(\br)$ to
changes in $\rho$ through the inverse capacitance matrix $W$: \begin{equation}
\delta\mathrm{V}(\br)=\int d^{2}r^{\prime}W\left(\br,\br^{\prime}\right)\delta\rho\left(\br^{\prime}\right).\label{eq:ES}\end{equation}
 If a time-independent steady state is reached, then we have simply
$\nabla\cdot\bj=0$, and the precise form of $W$ is unimportant,
but the form of $W$ will be relevant for time-dependent solutions.

In a Corbino geometry, one specifies the potential on the inner and
outer boundaries of the sample, and one looks for a solution for $V(\br)$
consistent with these boundary conditions. If one assumes $\sigma_{H}$
to be a constant, the Hall current cannot contribute to $\nabla\cdot\bj$
in the interior of the sample, so it does not appear in Kirchoff's
equations. Consequently, the solution $V(\br)$ is independent of
$\sigma_{H}$ and we may, for simplicity set $\sigma_{H}=0$. To recover
the Hall current, one simply inserts the resulting solution for $\be$
into the second term in (\ref{eq:curr}), at the end of the calculation.

Condition (\ref{eq:jdiss}) on $\bj^{d}$ allows us to define a scalar
\textit{Lyapunov functional} as

\begin{eqnarray}
G[\mathrm{V}] & = & \int d^{2}r\left[g(\be)+\frac{\lambda}{2}\left(\nabla\cdot\be\right)^{2}\right]\label{eq:G-def}\\
g & \equiv & \int_{0}^{\be(\br)}d\be^{\prime}\cdot\bj^{d}(\be^{\prime})\end{eqnarray}
 A variation of (\ref{eq:G-def}) is given by \begin{equation}
\delta G=\int d^{2}r\nabla\cdot\bj^{l}\delta V-\int_{bound}ds\,\hat{\mathbf{n}}\cdot\bj^{l}\delta V,\label{eq:dG}\end{equation}
 where $\bj^{l}=\bj^{d}-\lambda\nabla^{2}\be$. The second integral
vanishes on equipotential boundaries, or in the absence of external
currents. Then, if $\sigma_{H}$ is independent of position, the extrema
of $G$ are found to be steady states, with $\nabla\cdot\bj=0$. Using
the positivity of the inverse capacitance matrix $W$, one may show
that $G[V(t)]$ is indeed a Lyapunov functional, i.e. a non-increasing
function of time, so that its minima are stable steady states. In
general, $G$ may have multiple minima. Any initial choice of $V(\br)$
will relax to some local minimum of $G$, though not necessarily the
{}``ground state'' with lowest $G$. Nevertheless, one might expect
that in the presence of noise, the system might tend to escape from
high-lying minima and wind up in a state with $G$ close to the absolute
minimum.

We shall assume here that the function $g$ depends only on the norm
squared $E^{2}$ of the electric field, as would be appropriate in
the limit of a uniform isotropic electron system, assuming that there
is no unique axis picked out by external factors such as the polarization
of the microwave field. By hypothesis, under conditions where ZRS
occurs, the function $g$ must have its absolute minimum at a non-zero
value of the electric field, $E=E_{c}$. Expanding about this minimum,
and keeping only the nontrivial terms of lowest order, we may write\begin{equation}
g(\be)=g(E_{c})+\frac{\sigma_{c}}{8E_{c}^{2}}\left(E^{2}-E_{c}^{2}\right)^{2}\label{eq:gquad}\end{equation}
 where the coefficient $\sigma_{c}$ has the dimensions of a conductivity.
In the absence of other information, we take $\sigma_{c}$ to be of
the order of the dark conductivity (see Sec. \ref{sec:Experimental-Prospects}
for estimates). When $E=E_{c}$, the dissipative current, $\bj^{d}=\partial g/\partial\be$,
will vanish, and $\sigma_{\alpha\beta}^{d}(\be)=\sigma_{c}E_{\alpha}E_{\beta}/E^{2}$.

In the following calculations, for reasons of simplicity, and also
to avoid introduction of additional parameters, we shall assume that
Eq. (\ref{eq:gquad}) is exact for all values of $E$. (Actually,
only the range $0<E<E_{c}$ is important for our calculations.) We
shall also assume that $E_{c}$ and $\sigma_{c}$ are independent
of position.

The charge density $\rho(\br)$ is naturally broken into two parts,
\begin{equation}
\rho(\br)=\rho_{0}(\br)+\delta\rho(\br),\end{equation}
 where $\rho_{0}(\br)$ is the density in thermal equilibrium with
no incident microwave radiation (by definition, the electrochemical
potential is constant everywhere under such conditions). In a Corbino
setup, for example, $\rho_{0}(\br)$ is the equilibrium density when
the contacts at both edges are set to $V=0$. In addition to having
variations due to a non-uniform local dopant density, $\rho_{0}(\br)$
could be tuned by a voltage on an external gate that is displaced
from the 2DEG by a distance which varies from one point to another.
The remaining term, $\delta\rho(\br)$, is the non-equilibrium charge
density produced by microwave-induced domain structure, and it is
this contribution which is responsible for the position dependent
electrochemical potential, according to Eq.~(\ref{eq:ES}).

In our calculations, we shall assume a local form for the capacitance
matrix $W$, so that the electrochemical potential $V(\br)$ and the
nonequilbrium charge density $\delta\rho(\br)$ at a given point are
simply proportional to each other: \begin{equation}
V(\br)=C^{-1}\delta\rho(\br)\label{eq:vcrho}\end{equation}
 This form will be correct if there is a parallel conducting gate,
set back from the 2DEG by a distance $d$ small compared to the typical
domain size, in which case we have $C\approx\ \epsilon\epsilon_{0}/d$,
where $\epsilon$ is the dielectric constant of the material between
the 2DEG and the gate. If a nearby conducting plane is absent, the
local capacitance model will not be strictly correct, but we would
expect to obtain qualitatively correct results by using (\ref{eq:vcrho}),
replacing the setback distance $d$ by a characteristic domain size.

We shall assume that the Hall conductance at point $\br$ is determined
in the usual way by the charge density $\rho$ at that point : \[
\sigma_{H}(\br)=\frac{\rho(r)}{B}\]
 Variations in the Hall conductance due to variations in $\rho_{0}(\br)$
will be crucial for the effects we investigate below. On the other
hand, variations in $\sigma_{H}$ due to variations in $\delta\rho(\br)$
play no role in the dynamics when one assumes a local capacitance,
as in Eq. (\ref{eq:vcrho}). The contribution of this term to $\nabla\sigma_{H}$
is parallel to $\nabla V$, and, therefore, gives no contribution
to the divergence of the Hall current. Hence, we shall neglect this
contribution and write the Hall conductance as \[
\sigma_{H}(\br)=\frac{\rho_{0}(\br)}{B},\]
 which is a quantity fixed at the outset, independent of the non-equilibrium
charge-density induced by the microwave radiation.

We finally note that the Lyapunov functional of Eq. \ref{eq:G-def}
misses one important piece of physics--while it incorporates the fact
that there is a high penalty for the field to change on length scales
shorter than the domain wall thickness $\sqrt{\lambda/\sigma_{c}}$,
it neglects the fact that, at least in the local capacitance model,
the domains cannot be arbitrarily big. More precisely, if we imagine
drawing lines through the domain that are parallel to the field inside
of it, then these line segments inside the domain cannot get too long.
Otherwise, the induced voltage difference between points at opposite
ends of these lines would result in an induced charge density $\delta\rho$
that alters the total electronic charge density by a large amount,
which should be taken into account in the functional form of the Lyapunov
function $g(\be)$. For example, one might need to take into account
variations in the parameters $E_{c}$ and $\sigma_{c}$ in Eq. (\ref{eq:gquad})
due to variations in $\delta\rho$. In a local capacitance model,
the quantity \begin{equation}
L_{\mathrm{m}ax}=\frac{\bar{\rho}_{0}}{C\, E_{c}},\label{eq:Lmax}\end{equation}
 where $\bar{\rho}_{0}$ is the mean electronic charge density in
the 2DEG, will clearly provide an upper cutoff to the size of any
domain. Rather than modifying the functional, we shall assume here
that the domains we consider are below such cutoff. Note that the
maximum domain size will tend to be largest if the setback distance
$d$ of the screening gate is large, so that the capacitance per unit
area is small.

\subsection*{Dimensionless parameters}

In our calculations below, we shall assume a rectangular sample with
linear dimensions $L_{x}=L$ and $L_{y}=\kappa L$. As postulated
above, in the domains phase, the magnitude of electric field away
from the domain walls be equal to $E_{c}$. We will then work with
dimensionless units, with distances measured in units of $L$, electric
field $\mathbf{E}(\mathbf{r})=-\nabla V(\mathbf{r})$ measured in
units of $E_{c}$, electrochemical potential $V(\mathbf{r})$ measured
in units of $E_{c}L$, conductivities measured in terms of the coefficient
$\sigma_{c}$, and, finally, time measured in units of $CL^{2}/\sigma_{c}$.

The dimensionless Lyapunov density, c.f. Eq. (\ref{eq:G-def}), we
consider is then $g(E)=(-\frac{E^{2}}{4}+\frac{E^{4}}{8})$. When
it is minimized, the electric field is fixed at 1, that is $E_{c},$
away from the domain walls. Including the nonlocal contribution giving
nonzero thickness to domain walls, the dimensionless longitudinal
current density implied by the Lyapunov form we use is

\begin{equation}
\bj^{l}(E)=(E^{2}-1)\frac{\be}{2}-\frac{l_{dw}^{2}}{4}\nabla^{2}\be,\label{eq:long current}\end{equation}
 where $l_{dw}$ is related to parameter $\lambda$ from Eq. \ref{eq:G-def}
via $l_{dw}=\frac{2}{L}\sqrt{\frac{\lambda}{\sigma_{c}}}.$

In the reduced units introduced above, we will be solving $\frac{\partial V}{\partial t}=-\nabla\cdot(\bj^{l}+\bj^{H})$,
for electrochemical potential $V(\br;\, t)$, on a $[0,1]\times[0,\kappa]$
rectangle. We have looked at two different sets of boundary conditions.
In the first one, corresponding to Corbino geometry, we assume periodic
boundary along $x$, but the potential is being fixed to be zero at
the top ($y=\kappa$) and the bottom ($y=0$) boundaries. In the second
case, corresponding to a torus, the periodic boundary conditions along
both directions are assumed. For simplicity, will only consider unidirectional
Hall conductivity $\sigma_{H}(y)$.

\section{\label{sec:Uniform-Hall-Conductance}Uniform Hall Conductance }

\begin{figure}[t]
\includegraphics[scale=0.4]{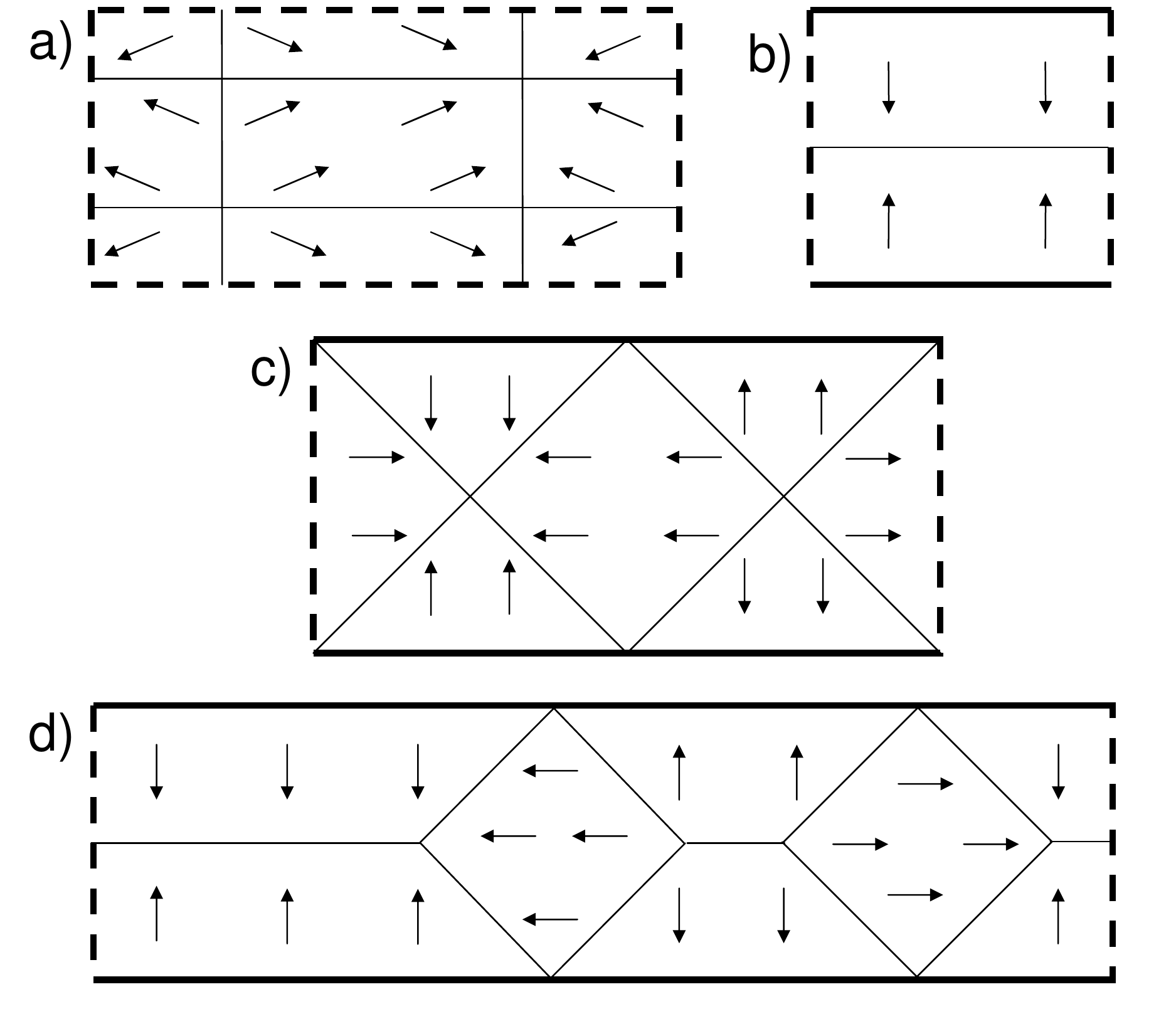}

\caption{\label{fig:unifhall} Schematics of domain patterns in a rectangular
sample with aspect ratio $\frac{L_{y}}{L_{x}}=\kappa$ for the case
of uniform Hall conductivity. Thick lines are sample boundaries, with
solid ones corresponding to boundaries where the electrochemical potential
is fixed, and the dashed ones corresponding to boundaries where periodic
boundary conditinos are employed. Thin lines are domain walls, and
the arrows indicate the direction of the electric field within the
domains. The magnitude of the field is constant away from the domain
walls. In panel a), the periodic boundary conditions along both directions
are assumed, and the shown state of the lowest Lyapunov energy has
two horizontal and two vertical domain walls. Due to the periodic
boundary conditions, this four-wall state is equal in energy to any
state obtained by translating the original one along either x or y
direction (or both). The rest of the panels correspond to the case
of periodic boundary conditions along x, with both the top and the
bottom boundaries being grounded. Panel b) corresponds to the case
of $\kappa>.5$, panel c) shows one of the degenerate states for $\kappa=.5$,
and panel d) represents one of a host of different possibilities arising
in the case of $\kappa<.5$.}
\end{figure}

We first consider the case where the equilibrium charge density, and,
therefore, Hall conductivity $\sigma_{H}$ is uniform. Then the domain
patterns observed should follow directly from Lyapunov energetics.
If $\gamma$ is the angle that the electric field forms with the domain
wall, then, for the Lyapunov density that we have assumed, the cost
of domain wall per unit length is proportional to $\sin^{3}\gamma$
(See the Appendix for the proof). If we \textbf{neglect} the energy
cost associated with wall crossings and exponentially small interactions
between well-separated walls, then we can describe the ground states,
that is the lowest Lyapunov energy states, of the system with the
two sets of boundary conditions introduced above. They are illustrated
in Fig. \ref{fig:unifhall} 

In the simpler case of all-periodic boundary conditions, the ground
state, shown in panel a) of Fig. \ref{fig:unifhall}, would have two
horizontal and two vertical walls, with electric field forming angle
$\delta$ ($\tan\delta=\kappa$) with horizontal walls. Of course,
there will be a degeneracy of these states due to translational invariance;
the pattern can be displaced by a arbitrary amounts in the horizontal
and vertical directions with no change in Lyapunov energy.

The situation is more complicated in the case of the Corbino-like
geometry, where we assume periodic boundary conditions in the $x$-direction,
but the top and bottom boundaries are grounded. Results actually depend
on the sample aspect ratio $\kappa$. For $\kappa>0.5$, the case
illustrated in panel b) of Fig. \ref{fig:unifhall}, the lowest energy
state has a single horizontal domain wall at $y=\kappa/2$. (There
is actually a doublet of ground states, related to each other by a
global flip of the signs of the electric field.) For $\kappa=0.5$,
there are two degenerate ground states, the first again consisting
of the single horizontal wall centered at the middle of the domain,
with the second, shown in panel c) of Fig. \ref{fig:unifhall}, having
four domain walls forming an angle of $\frac{\pi}{4}$ with the top/bottom
boundaries, and the electric field pointing along $\pm\hat{y}$ in
the domains adjacent to the top and bottom boundaries, and along $\pm\hat{x}$
in the domains that are fully in the bulk. This second solution can,
of course, be displaced in the horizontal direction by an arbitrary
amount, without penalty. For the case of $\kappa<0.5$, there are
an uncountable infinity of ground states, which may be thought of
as the combination of the single horizontal wall and the diagonal
wall states, as illustrated schematically in panel d) of Fig. \ref{fig:unifhall}.
The interactions between the walls will most likely break this continuum
down to a finite set of states, but such investigation lies outside
the scope of this paper.

It should be emphasized that the existence of non-equivalent degenerate
solutions in the case of $\kappa=0.5$, and the multiple degenerate
solutions for $\kappa<0.5$ are a peculiar feature of the quartic
Lyaponuv function (\ref{eq:gquad}) that we have assumed in our analysis.
In this case the Lyapunov energy per unit length of a domain wall
with field-angle $\gamma=\pi/4$ is smaller than that of a domain
wall with $\gamma=\pi/2$ by precisely a factor of $2\sqrt{2}$, just
compensating for the difference in total domain-wall length in the
two solutions. We believe that a more realistic assumption for the
form of the Lyapunov function $g$ would most likely favor the solution
with diagonal $\pi/4$ walls. (See discussion in the Appendix.) Similarly,
we expect that a more realistic Lyapunov function would favor a maximum
number of diagonal walls, and the shortest possible length of horizontal
wall, for aspect ratios $\kappa<0.5$. For any reasonable choice of
$g$, however, it is likely that multiple metastable configurations
will exist, for a range of values of $\kappa$, including configurations
with a single horizontal wall and configurations with two or more
diagonal walls.

\section{\label{sec:Exact-Solution-in}Corbino Geometry with a Gradient in
$\sigma_{H}$.}

A particularly simple model with a nonuniform equilibrium density
(and, hence, nonuniform Hall conductivity) will illustrate the possibility
of having a nonstationary state. Suppose that the Hall conductivity
varies linearly with position, $\sigma_{H}(y)=\alpha y$. This, of
course, makes sense in the context of a Corbino geometry, but not
on a torus. With linearly varying $\sigma_{H}(y)$, we find \begin{equation}
\mathrm{div}\,\bj^{H}=\sigma_{H}^{\prime}(y)\, E_{x}(\br)=\alpha E_{x}(\br).\end{equation}
 Suppose that $V_{0}(\br)$ satisfies $\mathrm{div\,}\bj^{d}=0$.
Then $V(\br,t)=V_{0}(\br+\alpha t\,\hat{\mathbf{x}})$ trivially satisfies
$\frac{\partial V}{\partial t}=-\mathrm{div}(\mathbf{j}^{d}+\mathbf{j}^{H})=-\alpha E_{x}(\mathbf{r})$.
In absolute units, this corresponds to a {\em drift velocity} \begin{equation}
\mathbf{v}=\hat{z}\times\nabla\sigma_{H}/C\,.\label{vdrift}\end{equation}
 Of course, this time dependence is not observable for a translationally
invariant state, as in the case of a single horizontal wall in the
center of the sample. However, if diagonal domain walls are present
in the sample, then, if one were to detect electron density at a single
point with, say, an SET, \cite{SET1,SET2} then one would find it
changing with time.

The question of whether the states other than the one with a single
horizontal wall would be present in an actual sample is hard to answer
definitively. Given a random initial configuration for $V(\br)$,
whether time evolution would take it to a state having diagonal walls
is a matter of the relatives size of the domains of attraction for
the such states compared with the domain of attraction for the state
with one horizontal wall. Time-evolving the continuity equation above
numerically (more details on the numerics are provided later in the
paper), we found that, for $\kappa\le0.5$, starting with a random
initial guess for the electrostatic potential, it is very likely that
we end up in the state with diagonal walls. Moreover, as mentioned
above, if we were to assume a more realistic form for the Lyapunov
function $g$ than the quartic form (\ref{eq:gquad}), we believe
that solutions with diagonal domain walls would most likely have lower
Lyapunov energies than the solution with a single straight wall, in
which case their basins of attractions would presumably be further
enlarged.

We may also consider the situation where there is a voltage difference
$V_{y}$ between the top and bottom edges of the system. As long as
$V_{y}$ is small compared to $E_{c}L_{y}$, the voltage can be accommodated
by small displacements of the domain walls, giving a non-zero average
of $E_{y}$, without producing a net current in the $y$-direction.
This can occur for the moving domains that we find when $\sigma_{H}$
depends linearly on $y$ in the same manner as for the stationary
solutions appropriate to constant $\sigma_{H}$. Thus, the property
of zero dc conductance, as measured in the Corbino geometry, is preserved
for the time-dependent solutions.

\section{\label{sec:Torus Numerics}Dynamics on a torus: Numerical Results}

While not as readily experimentally relevant as nonstationary states
in a Corbino setup, it is nonetheless instructive to consider the
kinds of nonstationary states we can get on a torus. We believe that
the results might be applicable to cases where equilibrium density
has correlation length that is much smaller than the sample size,
and that, in such cases, one might get away with modeling a smaller
piece of sample with periodic boundary conditions, rather than all
of it with Corbino or Hall-bar boundary conditions.

We consider here a sample of size $L_{x}=L_{y}=1$, with periodic
boundary conditions. As described in Sec. \ref{sec:Uniform-Hall-Conductance},
the equilibrium state for the case of uniform Hall conductivity has
two horizontal and two vertical domain walls with field-angle $\gamma=\pi/4$.
In our analysis, we shall assume that the spatially-varying part of
the Hall conductivity has the simple form \textbf{$\sigma_{H}(\br)=\alpha\sin2\pi n_{H}y$},
where $n_{H}$ is a positive integer. Our interests lie in trying
to understand how the solutions depend on the non-uniformity parameter
$\alpha$, for different $n_{H}$ and domain wall thicknesses $l_{dw}$.

As we tune nonuniformity $\alpha$ away from 0, the Hall current develops
a nonzero divergence in the equilibrium state. We find that, so long
as the $\alpha$ is small enough, the four-wall solutions can accommodate
such nonuniformity by bending. We may imagine that the gradient of
the $\sigma_{H}$ leads to a horizontal {}``drag force\char`\"{},
proportional to $\hat{z}\times\nabla\sigma_{H}$, which can be counteracted,
for small $\alpha$, by the {}``restoring force\char`\"{} produced
by a distortion of the domain structure, away from the shape that
minimizes the Lyapunov functional. By contrast, for large values of
$\alpha$, we find only solutions which are translationally invariant
along the $x$-direction, and have have two or more horizontal domain
walls with $\gamma=\pi/2$. The divergence of Hall current trivially
vanishes for such states.

\begin{figure}[t]
\includegraphics[scale=0.5]{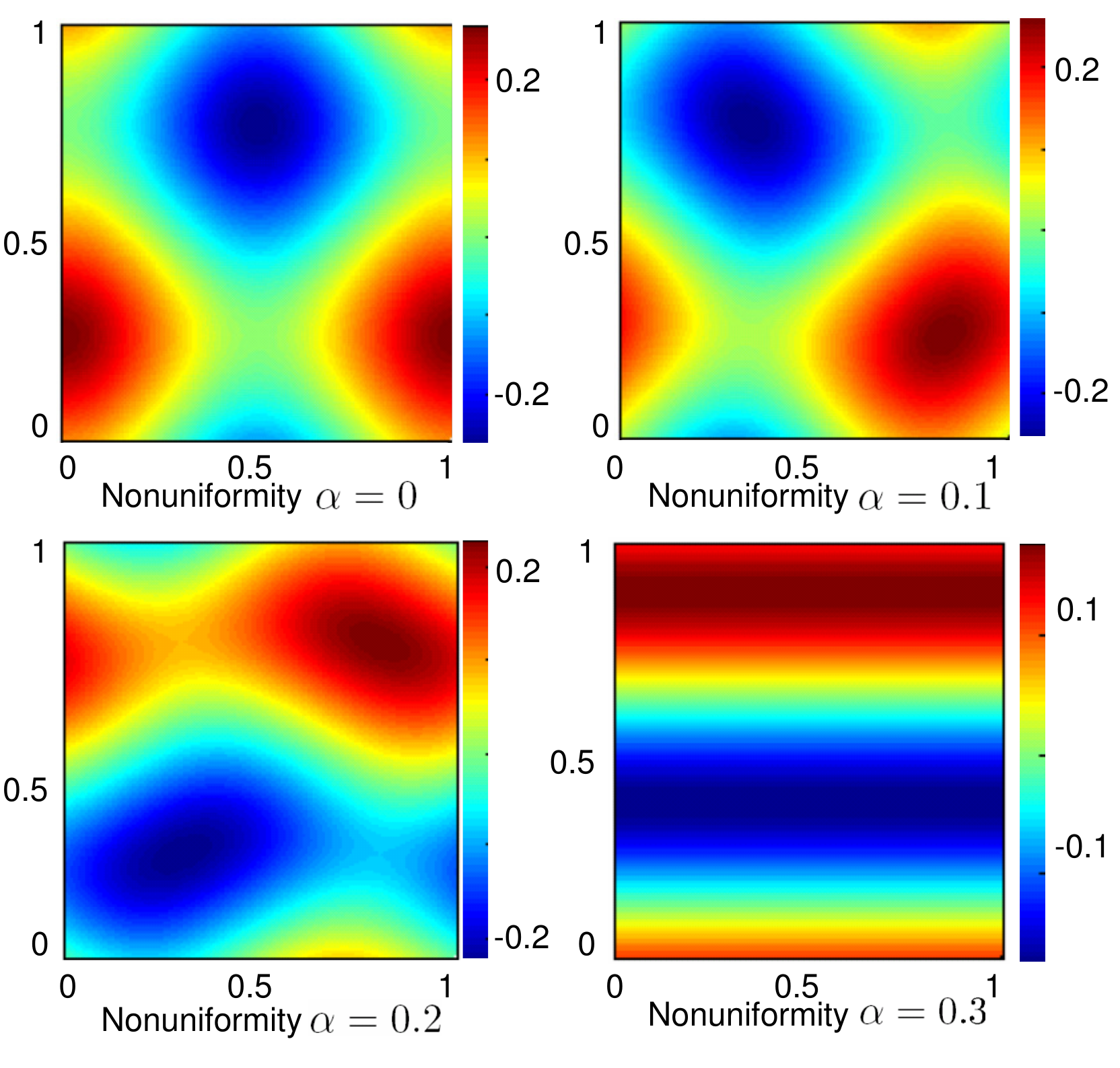}

\caption{\label{fig:ldw04} Numerically-obtained electochemical potential
$V(\br)$ for domain wall thickness $l_{dw}=.11$, $n_{H}=1$. All
$V(\br)$ shown are stationary. The transition between four-wall solution
at $\alpha=0$ and the one with two walls for $\alpha=0.3$ is continuous;
that is, it is not interrupted by the appearance of nonstationary
solutions at the intermediate values of $\alpha$.}
\end{figure}

What happens for intermediate values of $\alpha$ depends on $n_{H}$
and $l_{dw}.$ We find that if $l_{dw}$ is big enough, then one can
transition from the four-wall solution to the two-wall one without
ever encountering the nonstationary solutions. As seen in Fig.\ref{fig:ldw04},
the potential contour lines that have four-fold symmetry at $\alpha=0$
stretch and rotate as $\alpha$ is tuned up, until they become horizontal.

We find that increasing $n_{H}$ with everything else held fixed similarly
suppresses the appearance of nonstationary states. At $l_{dw}\lesssim.11$,
however, nonstationary states do exist. For example, when $l_{dw}=.028$
and $n_{H}=1$, we find nonstationary states for $\alpha\in[.07,.7]$.
Oftentimes, stationary and nonstationary states (perhaps even of different
types, see below) can exist at the same value of $\alpha$. Which
one we find in our simulations is both a matter of the initial conditions,
as well as the relative size of the basin of attraction for that particular
solution.

The nonstationary states we have seen can be broken into three groups.
The first is intimately connected to the solution one gets at small
$\alpha$. The periodic solutions of the first kind can be thought
as the system tunneling from one small $\alpha$ state to the one
connected to it by translation by $\frac{1}{2}\hat{x}$. Figure shows
snapshots of such a solution, plotting $V(\br,\, t_{i})$ for $t_{i}=i\frac{T}{8}$,
with $i=\{0,1,...,7\}$ and $T$ being the period, so that $V(\br,\, t+T)=V(\br,\, t).$
\begin{figure*}[t]
\includegraphics[width=180mm,keepaspectratio]{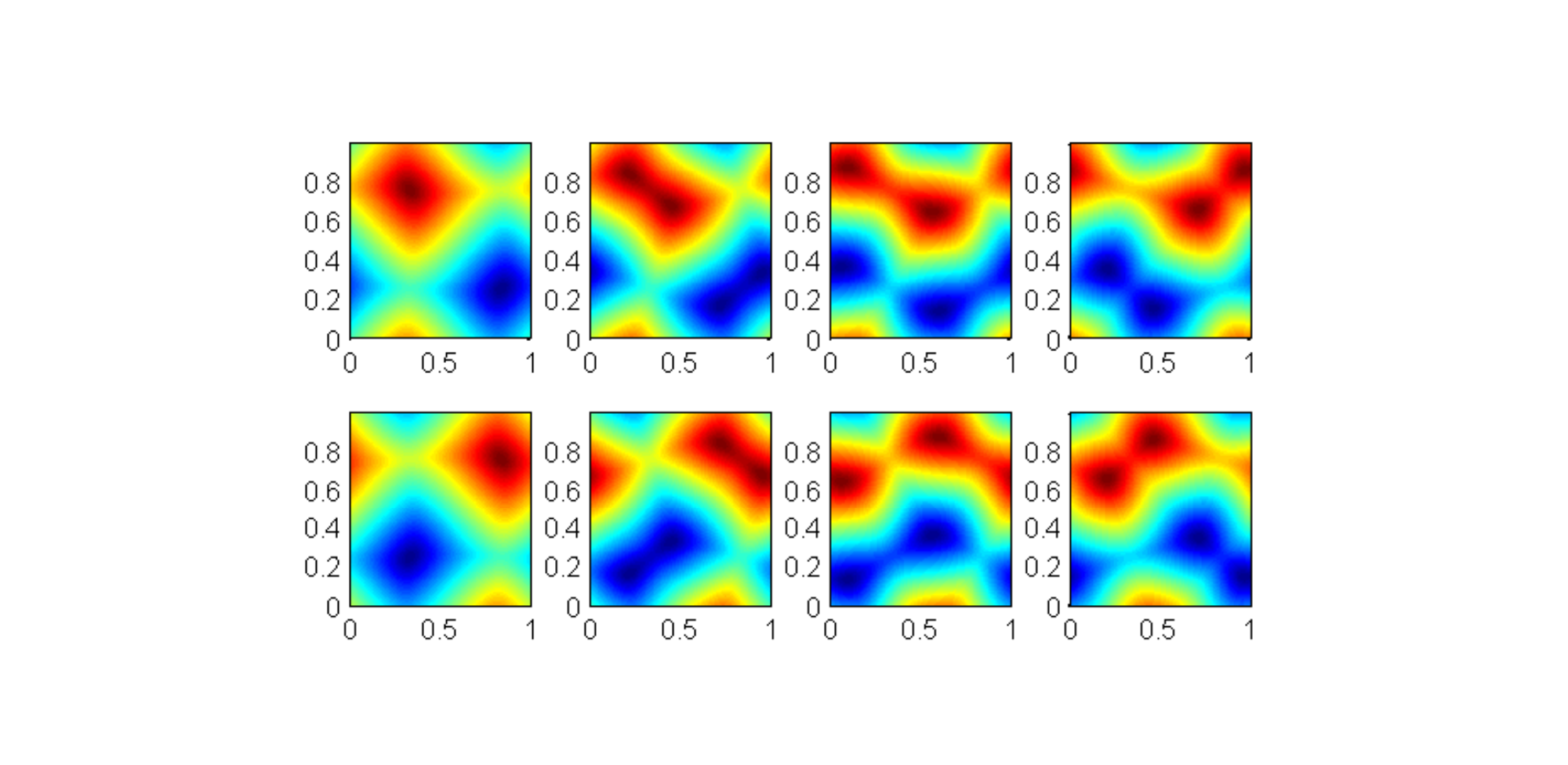}

\caption{\label{fig:type1} Numerically obtained electrochemical potential
$V(\br,\, t)$ for domain wall thickness $l_{dw}=.028$, $n_{H}=1$,
$\alpha=0.1$ at different points during its period. The solution
here {}``tunnels'' from one $\alpha=0$ solution to the one related
to it by translation by $\hat{x}/2$.}
\end{figure*}

In the periodic solutions of the second kind, the time-dependent potential
is confined to only half of the sample, and has the form $V(\mathbf{r}-\hat{x}vt)$,
representing a structure which drifts to the right or to the left
with a velocity $v$. This behavior is, of course, similar to what
we found in the Corbino geometry with a uniform gradient of $\sigma_{H}$.
Indeed, the observed values of $v$ are, to a good approximation,
related to the average value of $\nabla\sigma_{H}$ across a domain
in the same way as the drift velocity was related to the gradient
of $\sigma_{H}$ in Eq. (\ref{vdrift}) of Sec.~\ref{sec:Exact-Solution-in}.

Finally, the periodic solutions of the third type, seen only for $n_{H}=1$,
can be thought as the combination of two solutions of type II: in
one half of the sample, the potential moves to the right, while in
the other half it moves to the left, with more complicated processes
happening at the boundary between the two regions.

As examples of these behaviors (all but the first one not shown in
the figures), for $\alpha=.1$, $l_{dw}=.028$, we have found periodic
solutions for $n_{H}=1,\,2,\,3$, with the periods being 2.49 ($n_{H}=1$,
type I), 1.90 ($n_{H}=2$, type II), 1.64 ($n_{H}=3$, type I), and
1.21 ($n_{H}=3$, type II).

\section{\label{sec:Torus-Analytics}Analytic Results on a Torus}

We would like to understand what controls the periods mentioned towards
the end of the previous section. The quartic Lyapunov density of Eq.
\ref{eq:gquad} is assumed throughout this section. It turns out to
be useful to analyze the stability of the simplest solution, the one
with two domain walls perpendicular to y-axis. With the walls fixed
at $y=\frac{1}{4},\,\frac{3}{4}$ (we keep the aspect ratio $\kappa$
at 1), and the field varying along y-axis only, the continuity equation
is satisfied for all values of nonuniformity $\alpha$, as well as
all values of $n_{H}$ and domain-wall thickness $l_{dw}$. However,
the numerical evidence cited above at the very least seems to suggest
that, as we decrease $\alpha,$ the basin of attraction for the two-wall
solution rapidly shrinks. We will demonstrate below that, below some
$\alpha_{c}(l_{dw}),$ the two wall solution becomes unstable. As
we have no analytic solution to the full-blown nonlinear partial-differential
equation, the technique of choice will be to linearize the differential
equation satisfied by the potential around the two-wall equilibrium
and see whether the eigenmodes grow or decay with time for various
values of $\alpha$ and domain wall thickness $l_{dw}$. After linearizing
the equation, we will separately consider the effects that nonzero
$l_{dw}$ and nonzero $\alpha$ have on the nature of eigenmodes and
the corresponding eigenvalues. We will then try to understand what
happens when both of the parameters are nonzero.

\subsection{Linearized Continuity Equation}

Linearizing around the two-wall solution, we see that changing the
field by $\delta\be$ changes the longitudinal current by

\[
\delta\bj^{l}=\frac{E_{0}^{2}-1}{2}\delta\mathbf{E}+E_{0}^{2}\cdot\delta E_{y}-\frac{l_{dw}^{2}}{4}\nabla^{2}\delta\be\]
 Here, $E_{0}(y)$ is the field corresponding to the solution with
two horizontal walls. For the quartic Lyapunov density, the form of
$E_{0}(y)$ is derved in the Appendix for the case of a single domain
wall. That result should be applicable in the current case of two
domain walls, so long as the separation between them greatly exceed
domain wall thickness, $l_{dw}\ll.5$. The change in the Hall current
is $\delta\bj^{H}=\sigma_{H}(-\delta E_{y}\hat{x}+\delta E_{x}\hat{y})$.
Employing $\delta\be=-\nabla\delta V$, we see that the corresponding
divergences turn out to be:\begin{eqnarray*}
\mathrm{div}\delta\bj^{l}& =&-\frac{E_{0}^{2}-1}{2}\nabla^{2}\delta V-\frac{3}{2}\,\partial_{y}\delta V\,\partial_{y}E_{0}^{2}\\
&&{} -E_{0}^{2}\,\partial_{yy}\delta V+\frac{l_{dw}^{2}}{4}\nabla^{4}\delta V.
\end{eqnarray*} Furthermore, since $\sigma_{H}$ depends only on $y$, we find $\mathrm{div}\,\delta\bj^{H}=(\partial_{x}\delta V)(\partial_{y}\sigma_{H})$.
The charge continuity equation, aided by the assumption of the local
capacitance model, then tells us that \begin{eqnarray*}
\partial_{t}\delta V&=&-\mathrm{div}\delta\bj= \frac{E_{0}^{2}-1}{2}\nabla^{2}\delta V+\frac{3}{2}\,\left(\partial_{y}\delta V\right)\,\partial_{y}E_{0}^{2}\\
&&+E_{0}^{2}\,\partial_{yy}\delta V-\frac{l_{dw}^{2}}{4}\nabla^{4}\delta V+\left(\partial_{y}\sigma_{H}\right)\partial_{x}\delta V.
\end{eqnarray*}

Different Fourier modes along x-axis will be decoupled, motivating
the ansatz $\delta V(x,y)=e^{at}e^{2\pi nxi}h(y)$. Our differential
equation then becomes:

\begin{align}
\left(a+\frac{q_{x}^{4}l_{dw}^{2}}{4}+q_{x}^{2}\frac{E_{0}^{2}-1}{2}-\left(q_{x}i\right)\sigma{}_{H}^{\prime}\right)\, h\,-\,\frac{3}{2}\,\partial_{y}E_{0}^{2}\, h^{\prime}\,\nonumber \\
-\,\left(\frac{3E_{0}^{2}-1}{2}+\frac{q_{x}^{2}l_{dw}^{2}}{2}\right)\,\sd{h}\,+\,\frac{l_{dw}^{2}}{4}\fd{h} & =0,\label{eq:eigenmode}\end{align}
 where $q_{x}=2\pi n$.

\subsection{Case of $\alpha=0$}

We first consider the case of uniform Hall conductivity but nonzero
domain-wall thickness. Equation (\ref{eq:eigenmode}) now turns into

\begin{align*}
\left(a+\frac{q_{x}^{4}l_{dw}^{2}}{4}+q_{x}^{2}\frac{E_{0}^{2}-1}{2}\right)\, h\,-\,\frac{3}{2}\,\partial_{y}E_{0}^{2}\, h^{\prime}\,\\
-\,\left(\frac{3E_{0}^{2}-1}{2}+\frac{q_{x}^{2}l_{dw}^{2}}{2}\right)\,\sd{h}\,+\,\frac{l_{dw}^{2}}{4}\fd{h} & =0\end{align*}
The regions immediately surrounding the domain wall and the regions
far away from it should be considered separately, turning this problem
into a WKB-type computation. The interval $[0,\,1]$ is broken into
four regions, two bulk domain regions and two thin (extending a few
$l_{dw}$) regions around the domain walls at $\frac{1}{4}$ and $\frac{3}{4}$.
The equation is separately solved in each region, then different solutions
are patched at various boundaries. We will assume a power series expansion
for eigenvalue $a$ and eigenfunction $h(y)$: $a=\sum a_{i}l_{dw}^{i},\,\,\, h(y)=\sum h_{i}(y)\, l_{dw}^{i}$.
Trivially, we have $a_{0}=0$ and $h_{0}=C$.

Note that $E_{0}^{2}$ is symmetric under reflection through either
the center of domain wall or the center of the domain. That means
that we can pick the solution to be symmetric or antisymmetric with
respect to reflection through the centers of domains or through the
domain walls. There are four cases to consider, and we will first
briefly mention those two where the solution is antisymmetric under
the reflection through domains' midpoints. Going order by order in
the expansion, one can show that all contributions are identically
zero. We now turn to the more interesting case of midpoint-symmetric
solutions.

Solving for $h_{i}$'s in the bulk and in the boundary layers, and
matching these solutions yields the terms in the eigenvalues expansion.
We note that since $h_{i}$ in the boundary layers are most conveniently
written as function of $z=(y-y_{wall})/l_{dw}$, the matching conditions
would be \begin{eqnarray*}
h_{i}^{bulk}(y_{wall}^{+}) & = & h_{i}^{bndry}(\infty)\\
\frac{dh_{i}^{bulk}(y)}{dy}\Bigr|_{y=y_{wall}^{+}} & = & \frac{dh_{i+1}^{bndry}(z)}{dz}\Bigr|_{z=\infty},\,\mathrm{etc.}\end{eqnarray*}
Assuming the quartic form (Eq. \ref{eq:gquad}) for the Lyapunov function
$g(\be)$, we have worked out the first few leading terms in the expansion
of $\sym$, the eigenvalue corresponding to the mode that is invariant
under reflection through both domains wall and domain midpoints (that
is, $h_{0}^{domain\,1}=h_{0}^{domain\,2}$) and $\asym$, the eigenvalue
corresponding to the mode that is invariant under reflection through
domain midpoints but acquires a minus sign when reflected though the
domain walls (that is, $h_{0}^{domain\,1}=-h_{0}^{domain\,2}$). As
the computation is relatively straightforward, we will omit the details
of it here and present the final result:\begin{align}
\sym & =2q_{x}^{2}l_{dw}-\frac{1}{6}q_{x}^{4}l_{dw}^{2}\nonumber \\
\asym & =-\frac{2}{3}q_{x}^{2}l_{dw}-\frac{13}{54}q_{x}^{4}l_{dw}^{2}\label{eq:expresults2}\end{align}

These expressions illustrate two points. As expected, for high-$q_{x}$
modes, both eigenvalues are negative, simply reflecting the fact that
modulation of potential on wavelengths smaller than domain wall thickness
are strongly suppressed. For relatively small $q_{x}$, though, $\sym>0$.
In particular, considering $q_{x}=2\pi$, time evolution of the eigenmode
(see lower right panel of Fig. \ref{fig:linearregime}) takes us from
the potential that has two horizontal walls to the solution that we
know to be stable, the one with two horizontal and two vertical walls,
i.e. panel a) of Fig. \ref{fig:unifhall}.

Finally, we emphasize that the periodic conditions along $y$ caused
the zeroth order (in $q_{x}^{2}l_{dw}$) term to vanish. In a Corbino
setup, the zeroth order term is negative, thus causing both eigenvalues
to be negative for thin enough domain walls. This implies that the
solution with two horizontal walls would be locally stable in the
Corbino geometry.

\subsection{Case of $l_{dw}=0$\label{sub:zerothickness}}

In the limit of zero domain wall thickness, $E_{0}^{2}=1$ everywhere
but on the domain walls, where $\partial_{y}E_{0}^{2}$ diverges.
Setting $l_{dw}=0$ in Eq. \ref{eq:eigenmode} must then be accompanied
by the prescription of how to deal with $h(y)$ on the domain walls.
This is not a straightforward task. Instead, we shall take a stab
by {\em{assuming}} a simple boundary condition, namely that $h^{\prime}=0$
as one approaches the domain walls. This boundary condition will suffice
to determine the eigenvalues, as (\ref{eq:eigenmode}) is now simplified,
away from the domain walls, to a second order differential equation,
\[
\left(\, a-2\pi q_{x}\alpha i\,\cos2\pi\left(y+\beta\right)\,\right)\cdot h=\sd{h}.\]
 (Since the positions of domain walls are fixed, we have introduced
an offset parameter $\beta$ to study the general case where the domain
wall is not necessarily located at an extremum of $\sigma_{H}$.)

We shall be interested in understanding the behavior of eigenvalue
$a(\alpha)$ for small $\alpha$; in particular, we will be interested
in the most physically relevant eigenvalue, with the largest (most
positive) real part, as $\alpha\rightarrow0$, $a(\alpha)\rightarrow0$.
Note that the requirement that $h^{\prime}$ vanishes at each domain
wall essentially decouples different domains. The goal is to work
out the expansion of eigenvalue $a$ and eigenfunction $h(y)$ in
powers of $\alpha$: $a=\sum\tilde{a}_{j}\alpha^{j}$ and $h(y)=\sum\tilde{h}_{j}(y)\alpha^{j}$.
One readily sees that real part of eigenvalue (and eigenfunction)
only contains even powers of $\alpha$, while the imaginary part has
only odd powers.

Starting with $\tilde{a}_{0}=0$ and $\tilde{h}_{0}(y)=1$, we can
get $\tilde{h}_{1}(y)$ by integrating twice the differential equation
it obeys, \[
\tilde{a}_{1}-2\pi q_{x}\,\cos\,2\pi(y+\beta)=\sd{\tilde{h}_{1}}.\]
 The assumed boundary condition, $\tilde{h}_{1}^{\prime}\left(\pm\frac{1}{4}\right)=0$,
then fixes $\tilde{a}_{1}$. In a similar manner, $\tilde{h}_{2}$,
$\tilde{a}_{2}$, and higher order terms in the expansion can be obtained.
The first few terms in the eigenvalues expansion are as follows: \begin{eqnarray*}
\tilde{a}_{1}&=&i\,4q_{x}\,\cos2\pi\beta;\nonumber \\
\tilde{a}_{2}&=&-\frac{q_{x}^{2}}{3}\left(2-\frac{12}{\pi^{2}}+\left(\frac{1}{2}-\frac{12}{\pi^{2}}\right)\,\cos4
\pi\beta\right);\nonumber \\
\tilde{a}_{3}&=&-i\,\frac{q_{x}^{3}\cos2\pi\beta}{45}\Bigg(\frac{1}{2}-\frac{120}{\pi^{2}}+\frac{1080}{\pi^{4}}+\\
&&+\left(\frac{1080}{\pi^{4}}-\frac{435}{4\pi^{2}}+\frac{1}{2}\right)\cos4\pi\beta\Bigg).\label{eq:expansionresults1}
\end{eqnarray*} The next term in the expansion would not be given explicitly here,
but we find it useful to give numerical values of the $a_{i}'s$ here:
in the case of $\beta=0,\, n=1$, one finds $\tilde{a}_{1}=25.13,\,\tilde{a}_{2}=-.8987,\,\tilde{a}_{3}=.0149,\,\tilde{a}_{4}=-.0041$.
We also find that the first two terms in the expansion of real part
of $a$, $\tilde{a}_{2}(\beta)\alpha^{2}+\tilde{a}_{4}(\beta)\alpha^{4}<0$
for all $\beta$ and for $\alpha$ smaller than about 8.36.

We finally comment that making shift $\beta\rightarrow\beta+\frac{1}{2}$
is equivalent to moving from one domain to the other. This leaves
the real part of eigenvalue unchanged while flipping the sign of the
imaginary part. It follows that eigenvalues come in conjugate pairs,
and that corresponding eigenfunctions must identically vanish in one
of the domains. This last result, of course, is a consequence of our
assumption that $h'=0$ at the domain walls, but it is reminiscent
of the type III periodic solutions in our numerical results, where
charge is moving in one directions in one of the domains, and in the
opposite direction in the other domain. In this light, it is not at
all surprising that the solutions of this kind show up at fairly large
$\alpha$, where the fact that the domain wall thickness is actually
nonzero is not qualitatively important. This gives us reassurance
that setting $h^{\prime}(y)=0$ on domain walls was a reasonable guess.

\subsection{Case of $l_{dw},\,\alpha\neq0$\label{sub:Nonzero ldw alpha}}

We are now ready to address the case where the Hall conductivity is
nonuniform and the domain walls have nonzero thickness. We restrict
the analysis to the cases where $\alpha$ is at most $O(1)$ and we
consider the mode with the lowest $q_{x}$, $q_{x}=2\pi$. The analysis
of the previous subsection then implies that, for $\alpha/l_{dw}\gg1$,
we might expect $\Re\, a(\alpha,l_{dw})<0$. The eigenmode then is
a decaying one, and we find that the solution with two horizontal
walls is stable. Below some $\alpha_{c}$ (which is a function of
$n_{H}$ and $l_{dw}$), however, the mode will start growing, and
we find that putting (just about any) perturbation on top of the two-wall
solution and time-evolving the resulting guess would take us away
from it. Indeed, for $\alpha/l_{dw}\ll1$, we expect $a(\alpha,l_{dw})\approx\sym(l_{dw})>0$
(see Eq. \ref{eq:expresults2}). 

There is, additionally, another special value of $\alpha$, $\alpha_{d}$,
below which the eigenvalues are real and unequal to each other, whereas
above it the eigenvalues are complex and are conjugate to each other.
At that special point, the eigenvalues are, of course, equal to each
other. We expect that $\alpha_{d}/l_{dw}$ is $O(1)$. We, however,
can't estimate $a(\alpha_{d},l_{dw})$; thus we can't establish whether
$\alpha_{c}>\alpha_{d}$ or the other way around. It is plausible,
though, that nonstationary solutions would exist for $\alpha\in[\alpha_{d},\alpha_{c}]$,
as, in that case, we would have oscillatory runaway modes.

As for the functional form of the eigenmodes, we expect them to be
mostly symmetric between two domains for $\alpha<\alpha_{d}$. If
our guess for the boundary conditions in the previous subsection were
a reasonable one, we would expect the modes to have support in one
of the domains only for $\alpha/l_{dw}\gg1$.

%It is important to note that, generally, there cannot be a general expansion of the type . This has to do with the fact the leading term in the expansion, i.e.  term, must be either symmetric or antisymmetric (with respect to reflection through a domain wall) in the case of finite domain wall thickness and uniform Hall conductivity, but it must vanish in one of the domains in the case of nonuniform Hall conductivity and vanishing domain wall thickness. [ISN'T THIS JUST THE RESULT OF OUR ASSUMPTION, WHICH WE HAVE NOT PROVED?] At this point, we have been unable to obtain the estimate of  that is consistent with our numerical findings.

\section{\label{sec:Simulation-Details}Simulation Details}

\subsection{Numerical Setup \& Methodology}

\begin{figure}[t]
\includegraphics[width=180dd,keepaspectratio]{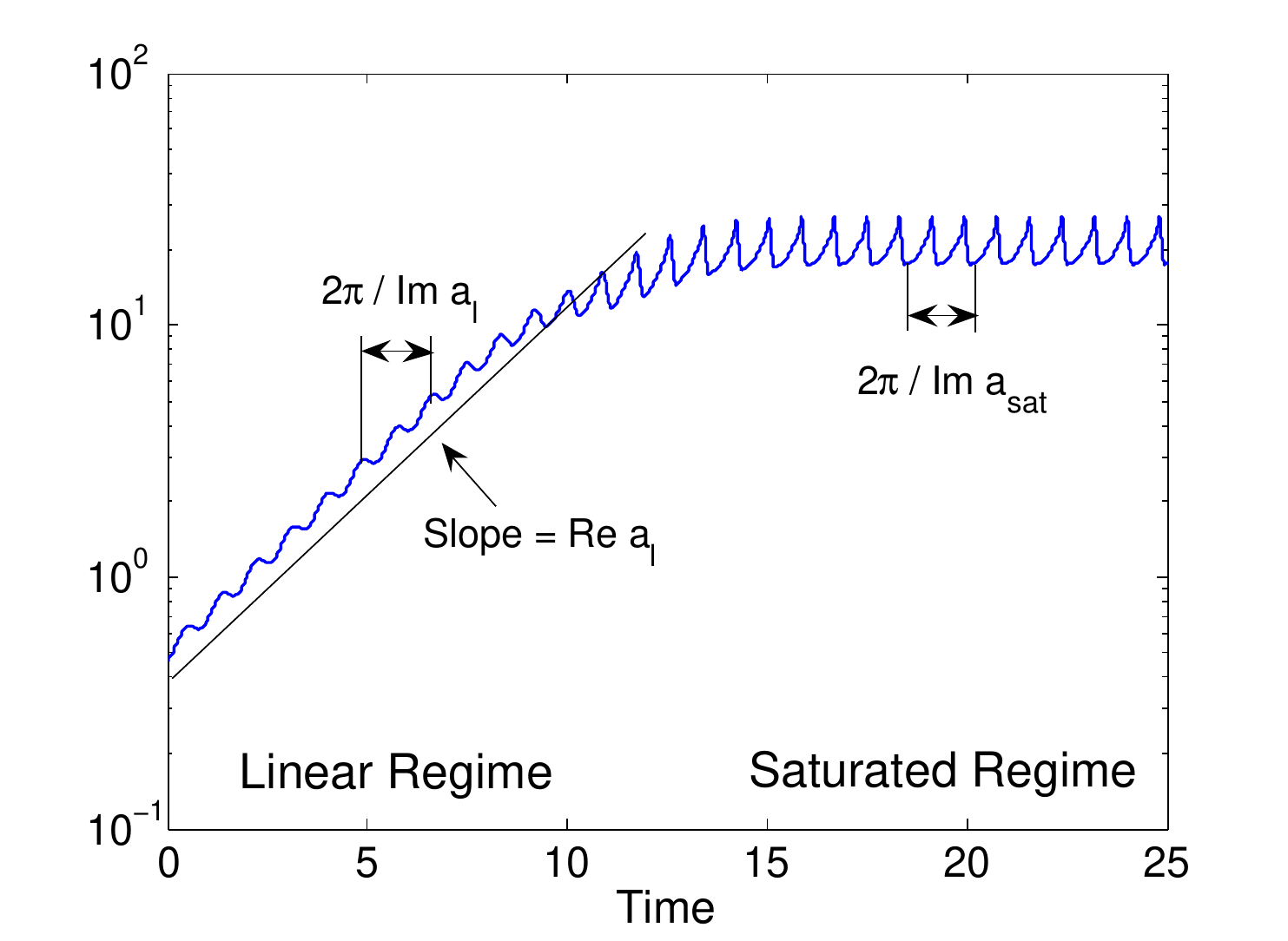}

\caption{\label{fig:time55y15f1} Time trace of $A\equiv\sqrt{\sum_{i}\left|\frac{\partial\rho_{i}}{\partial t}\right|^{2}}$,
with the sum being taken over grid points, can be used to extract
both the real and imaginary parts of frequencies. The transition from
linear regime (where the real part of the frequency $\omega_{\mathrm{l}}$
is nonzero) to the saturated one (where the real part of $\omega_{\mathrm{sat}}$
vanishes) is apparent.}
\end{figure}

The continuity equation was discretized on a triangular lattice of
grid points and then evolved in time. For the linear analysis part,
the initial guess consisted of a two-wall solution with perturbation
that only had one Fourier component in the direction parallel to the
walls (along x). Though the precise form of y-dependence of the perturbation
did not matter much, we have tried, among other things, a function
that is constant on both domains and one that vanishes in one domain
but is constant in the other one. In Fig. \ref{fig:time55y15f1},
we have plotted the root-mean-square amplitude $A(t)\equiv\sqrt{\sum_{i}\left|\frac{\partial\rho_{i}}{\partial t}\right|^{2}}$,
with the sum being taken over all grid points, as a function of $t$,
for a particular solution. As can be seen in the plot, for small values
of $t$ the amplitude increases exponentially over time, while also
exhibiting oscillations. From this portion of the plot, we can infer
both the real and imaginary parts of the eigenvalue $a_{l}$, which
governs behavior in the linear regime. At larger times, the exponential
growth has saturated, but oscillations persist. From this portion
of the plot, we can identify a second frequency, $a_{\mathrm{sat}}$
which characterizes oscillations in the saturated regime, as indicated
in the figure. %

When we were looking for various possible nonlinear solutions, the
initial guesses corresponded to the potential at each site being assigned
a random number between -0.5 and 0.5. Most of the analysis was carried
out on a lattice with mesh size $\frac{1}{55}$, though resolution
studies have also been carried out on grids with mesh about half as
big and quarter as big. Working with the rougher mesh has restricted
us to domain wall thicknesses $l_{dw}$ no smaller than about .03.

\subsection{Linear Regime Results}

\begin{figure}[t]
\includegraphics[width=80mm,keepaspectratio]{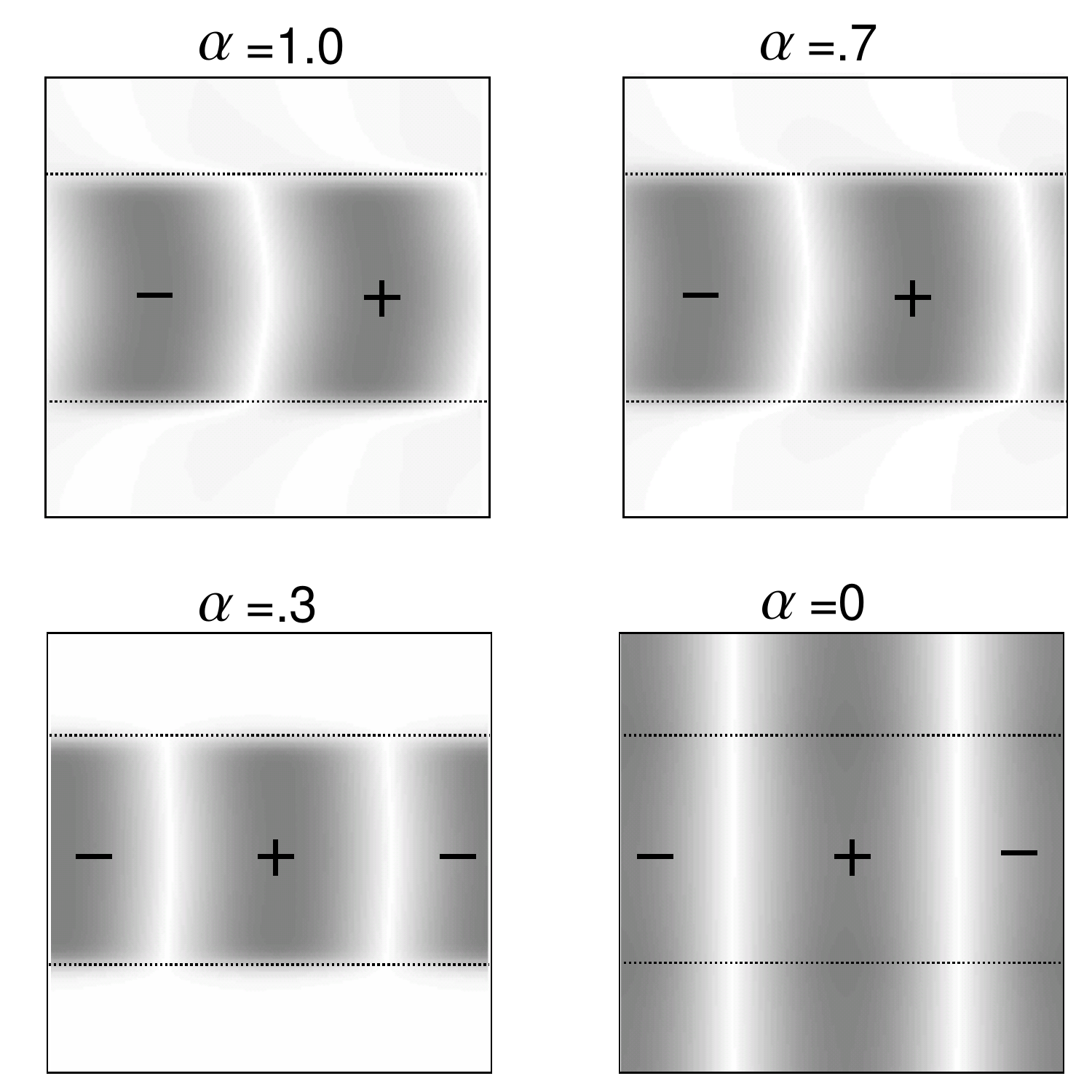}

\caption{\label{fig:linearregime} Plots of absolute value of eigenmodes of
the linearized continuity equation for all-periodic boundary conditions,
$l_{dw}=.028$, and $n_{H}=1$. The eigenmodes are deviations from
the solution with two horizontal walls; thin dashed lines indicate
the location of these domain walls. Plus and minus signs indicate
whether the deviation is positive or negative, while white color corresponds
to zero deviation. For sufficiently big values of the nonuniformity
$\alpha,$ all of the eigenmode's support is within one domain. On
the other hand, for $\alpha=0$, the (unstable) mode is symmetric
between two domains.}
\end{figure}

We first comment on the solutions and then on the eigenvalues. Recall
that the prediction was for eigenmodes to be largely localized to
one domain for large $\alpha$ and be symmetric between the two domains
around $\alpha=0$. And this is indeed what is seen in Fig. \ref{fig:linearregime}.%
The presence of free constants in the perturbative expansions we obtained
for the eigenmodes prevents us from making more detailed comparison
between theory and simulation.

\begin{figure*}[t]
\includegraphics[width=140mm,keepaspectratio]{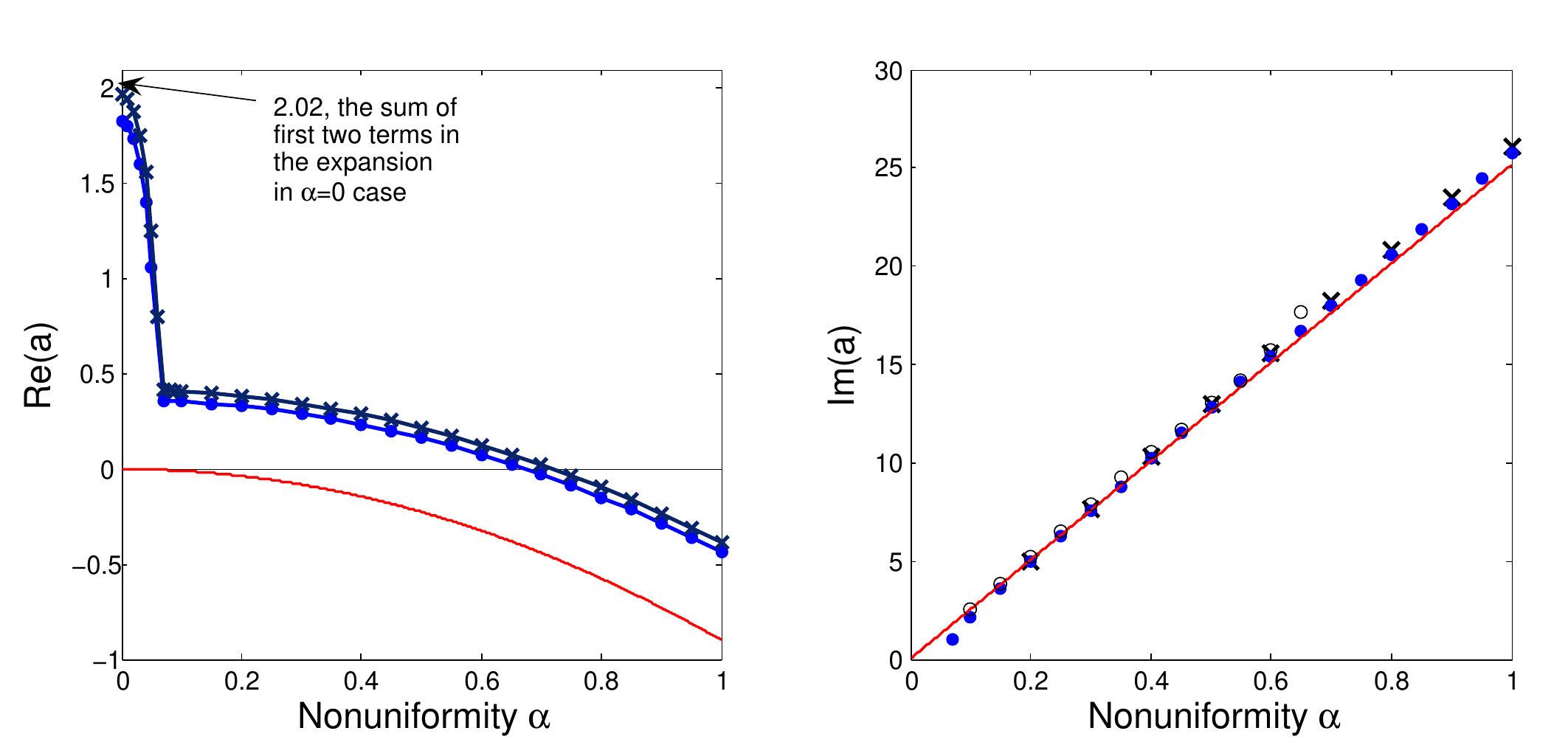}

\caption{\label{fig:omegas} Real and imaginary parts of eigenvalue $a$.
The left panel shows the real part of $a$ extracted from linear regime
simulations on rougher mesh (filled circles), finer mesh (crosses),
as well as the first term in perturbative expansion for the case of
$l_{dw}=0$ (solid line). The dashed lines connecting crosses and
filled circles are merely a guide for the eye. Right panel shows the
imaginary part extracted from simulations in the linear regime (with
filled circles corresponding to rougher mesh and crosses to finer
mesh), the nonlinear regime (open circles), and the first term in
the perturbative expansion for the case of $l_{dw}=0$ (solid line). }
\end{figure*}

The eigenvalues inferred from numerical simulations in the case of
$l_{dw}=.028,\, n_{H}=1$ are shown in Fig. \ref{fig:omegas}. For
$\alpha=0$ case, we get an excellent quantitative agreement between
the simulation results and the contributions of the first few terms
in perturbative expansion for the eigenvalue $a$, shown in Eq (\ref{eq:expresults2}).
As predicted, the eigenvalue remains real for a while before turning
complex, and these complex ones do indeed come in pairs. For this
particular domain wall thickness, we find, using the notation of Sec.
\ref{sub:Nonzero ldw alpha}, $\alpha_{d}\approx0.07,\,\alpha_{c}\approx0.7$.
We have found though, that if $l_{dw}$ is sufficiently big, the order
is reversed and one would have $\alpha_{d}>\alpha_{c}$. Note that
the curvature of $a(\alpha)$ for $\alpha>\alpha_{d}$ is quite close
to $\tilde{a}_{2}$ of Eq. \ref{eq:expansionresults1} (solid line
of Fig. \ref{fig:linearregime}). We can't predict the vertical offset
between the simulation results and the solid line--this is a direct
consequence of our inability to predict $\alpha_{d}$ and $a(\alpha_{d})$
which we have referred to earlier.

As for the imaginary part of the eigenvalue, the agreement between
the simulation results and the perturbative expansion is excellent.
It is noteworthy that not only is the agreement excellent in the linear
regime but even by the time we get to the saturated regime, the agreement
is still very good. Indeed, it was the this striking linear (in $\alpha$)
behavior exhibited by the full nonlinear solutions that has motivated
the whole expansion enterprise.

\section{\label{sec:Experimental-Prospects}Conclusions and Experimental Prospects}

We have seen that non-uniformities in the Hall conductance $\sigma_{H}$,
produced by inhomogeneities in the equilibrium electron density, can
lead, under certain circumstances, to time-dependent domain patterns
in the microwave-induced Zero-Resistance states seen in 2DEGs. Inhomogeneities
in the equilibrium density may result from disorder, such variations
in the density of ionized Si donors in the set-back doping layer,
or from density variations that are deliberately imposed on the sample.
Density variations of the latter type could be created by applying
voltages to one or more gates, located at appropriate distances from
the sample.

In our calculations, we have studied explicitly two simple models:
a Corbino-type geometry, with a linear dependence of $\sigma_{H}$
on the distance $y$ form one edge of the sample; and a model with
periodic boundary conditions and a variation in $\sigma_{H}$ that
has a sinusoidal dependence on one of the coordinates. In both cases,
we have found time-dependent solutions for appropriate choices of
the parameters.

The model with periodic boundary conditions may be useful as a first
step to examine the effects of random density fluctuations in a macroscopic
sample. In this case, we might identify the size $L$ of our model
with a region who size is given by the correlation length of the most
relevant density fluctuations in the physical sample. In our model
we found that time-periodic solutions occurred for density fluctuations
falling within a certain range. We speculate that similar time-varying
solutions may occur in the random system, with some regions of the
sample trying to generate time-dependent solutions of varying frequencies,
and other regions favoring decay to a time-independent solution. If
a time-dependent solution occurs in a macroscopic disordered sample,
presumably the power spectrum of the dynamic potential fluctuations
will contain many frequencies, and the domain-wall motion will be
chaotic. Since our starting equations are only valid on a length scale
greater than the cyclotron radius, of order 1 $\mu$m, we would not
want to consider intrinsic density fluctuations on the scale of the
set-back distance, but would restrict ourselves to longer wavelength
fluctuations of an extrinsic origin.

The Corbino model studied in Section \ref{sec:Exact-Solution-in}
suggests a way in which time-dependence may by implemented in a controlled
way, which may be most promising for experimental realization. Experimentally,
one might aim to create a circularly symmetric Corbino annulus, with
an electron density gradient along the radial direction. The density
gradient could be be produced by applying a voltage $V_{G}$ to a
front gate, whose distance $s$ from the 2DEG varied linearly with
the radial coordinate $r$. If $\epsilon_{s}$ is the dielectric constant
of the spacer between the gate and the 2DEG, then the imposed variation
in the equilibrium electron density would be given by \begin{equation}
\rho_{0}(r)-\bar{\rho}_{0}=\frac{V_{G}\epsilon_{s}\epsilon_{0}}{s(r)},\end{equation}
 where $\bar{\rho}_{0}$ is the mean charge density at $V_{G}=0$.
For example, if $s$ were to vary between 1 $\mu$m and 2 $\mu$m,
from the inner radius to the outer radius of the annulus, and if $\epsilon_{s}\approx13$,
one would need a gate voltage $V_{G}=0.09\mathrm{\, V}$ to produce
a density difference of $3.5\times10^{9}\mathrm{cm^{-2}}$ across
the sample, which would be 1\% of the average density for the sample
studied in Ref. \onlinecite{ZRS-exp2}. The gate, as well as the
spacer layer, should be transparent to the incident microwaves; this
might be achieved, for example, if the gate is made from a thin film
with high resistivity. Alternatively, the density gradient in a 2DEG
could be created, even if the spacer thickness is uniform by depositing
several separately-contacted gates on top of the spacer and putting
different voltages on them.

We can use the above numbers to estimate the characteristic frequency
of the charge oscillations that might be observed. Let us consider
a Corbino sample, whose width $w$ is much smaller than its circumference
$L$, say .1 mm and 10 mm respectively. In this regime, the curvature
of the annulus is not important, and the periodic states should resemble
those described in Sec. \ref{sec:Exact-Solution-in}. To apply the
calculations of Sec. \ref{sec:Exact-Solution-in}, we estimate the
capacitance per unit area $C=\epsilon\epsilon_{0}/d$ by equating
the distance $d$ to 1.5 $\mu$m, the average distance to the gate,
and use $\epsilon_{s}=13$ for the dielectric constant. If we assume
that the electron density changes by a factor $1+\beta$ across the
width $w$ of the annulus, then the drift velocity $\mathbf{v}$ in
Eq. (\ref{vdrift}) has magnitude $v=\beta\sigma_{xy}/wC$. If we
assume that the domains have a typical length of order $w$, then
the moving domain structure will lead to a time-dependence of the
charge density, at any given point, with a characteristic period given
by \begin{equation}
\tau_{p}=\frac{w}{v}=\frac{Cw^{2}}{\beta\sigma_{xy}},\end{equation}
 For the parameters considered here, if $\beta$ corresponds to a
density change of 1\%, we find $\tau_{p}\approx$ 15 ns.

As we have assumed that the domain size is set by the width $w$ of
the Corbino sample, to be consistent we should check that $w=100\,\mu$m
does not exceed the cutoff scale $L_{max}=\bar{\rho}/CE_{c}$ of Eq.
\ref{eq:Lmax}. There are a few estimates of $E_{c}$ in the theoretical
literature\cite{Assa,ZRS-DF2}; for the typical experimental density
$n=3\times10^{11}cm^{-2}$, the highest estimate would be on the order
20 V/cm. Experimentally, while $E_{c}$ has not been measured, some
of the results of Refs. \onlinecite{ZRS-exp3,ZRS-exp4} can be interpreted
to imply $E_{c}\sim$0.1 V/cm. Cautiously, using the higher estimate
for the critical field and the same estimate for capacitance per unit
area $C$ as above, we arrive at $L_{max}\approx$0.6 cm. So, in a
Corbino sample that is $100\,\mu$m wide, the size of the domain should
indeed be set by the the Corbino ring's width. By varying the sample
width, the spacer thickness, or the density gradient the time scale
of interest, $\tau_{p}$, can be tuned over a wide range. 

For the sample size we are considering here, we need to detect density
changes at frequency of about 70 MHz. This might be accomplished in
number of ways. An apporpriately tuned RF-SET, positioned on the surface
of the the sample, underneath the gate should be able to pick up the
fluctuations at that frequency. Alternatively, a small capacitor,
placed at the surface, might be able to pick up changes in density.
Finally, if one could put down the Ohmic contacts in the middlle of
the sample without destroying Zero-Resistance state, then monitoring
the voltage difference between the inside of the sample and its boundary,
held at constant potential, might show the time dependence we are
predicting. The magnitude of voltage fluctuations produced by the
domain of width $w$ is given by $E_{c}w$. For a domain of width
$w=.1$mm and $E_{c}=.1$V/cm, this would imply voltage varying by
1mV.

The parameter $\sigma_{c}$ in the Lyapunov function $g(\be)$ does
not enter directly in our estimate for the oscillation period in the
Corbino geometry. However, it should play a role in a system with
disorder, as it did in our model with periodic boundary conditions,
where the non-uniform drift velocity $v$ is in competition with the
tendency of the Lyapunov function to favor an incompatible domain
structure. The dimensionless parameter $\alpha$, which controlled
the transition from time-independent to time-dependent solutions in
Sec. \ref{sec:Torus Numerics}, may be written as $\alpha=\beta\sigma_{xy}/\sigma_{c}$,
where $\beta$ is here the fractional change in the electron density
between its minimum and maximum values. The value of $\sigma_{c}$
depends on details of the microscopic model, but it is expected generally
to be of the order of the longitudinal conductivity $\sigma_{xx}$
in the absence of microwave radiation. For the sample described in
Ref. \onlinecite{ZRS-exp2}, in a field of $\sim$.11T, where the
most prominent ZRS was seen once the sample was subjected to microwaves,
we estimate a value for $\sigma_{xx}$ of order $5\times10^{-5}$S,
while $\sigma_{xy}\approx5\times10^{-3}$S.

Finally, we note that our model for the Corbino geometry includes
an intentional density gradient, but ignores effects of random density
variations that may occur in an actual sample. If the random variations
are too large, it is possible that they would pin the domain walls
at specific locations in the sample, and defeat the tendency for motion
produced by the intentional gradient. At the present time, we have
little knowledge about the magnitude of density variations over the
range of length scales of interest and we cannot reliably estimate
the importance of any such pinning effects. It seems likely that pinning
effects could be overcome by employing intentional density gradients
that are sufficiently large.

\subsection*{Acknowledgments}

The authors are grateful for discussions with Assa Auerbach, Jacob
Krich, Ari Turner, and Amir Yacoby. This work was supported in part
by NSF grants DMR-0541988 and PHY-0646094.

\appendix
%dummy comment inserted by tex2lyx to ensure that this paragraph is not empty

\section{Alternative forms for the Lyapunov function.}

The Lyapunov energy cost of a domain wall depends on the form of the
Lyapunov function $g(\be)$, for $E<E_{c}$, as well as on the angle
$\gamma$ between the domain wall and the field $\be$ just inside
the domain on each side. (The angle must be the same on both sides
of the wall.) For the maximum angle $\gamma=\pi/2$, the field inside
the domain wall passes through zero at the center of the domain wall,
so that the entire range $0<E<E_{c}$ is reached within the wall.
If the field is not perpendicular to the wall, however, the component
parallel to the wall remains constant and non-zero throughout the
wall, and only the perpendicular component of $\be$ changes sign.
The domain wall is then determined by the form of $g$ in the range
$E_{c}\cos\gamma<E<E_{c}$.

In our calculations, we have used Lyapunov energy density (\ref{eq:gquad}),
which is a quartic function of the field strength $E$. It is the
simplest possible function that has the following desired properties:
it is an analytic function of $E^{2}$, and hence an an isotropic
analytic function of $\be$; it has an absolute minimum at a non-zero
value of $E$, and a local maximum at $E=0$. The function is then
determined by the position $E_{c}$ of the minimum and the curvature
$\sigma_{c}$ at $E=E_{c}$. However, if the quartic function is extended
to values of $E$ much larger than $E_{c}$, the longitudinal differential
conductivity, $d^{2}g/dE^{2}$ is seen to increase without limit,
going eventually as $E^{2}$, which does not seem reasonable. It might
be more reasonable to assume that at large values of $E$, the longitudinal
differential conductivity should approach a constant value $\sigma_{\infty}$.
Although we are not directly interested in the behavior of $g(E)$
for $E>E_{c}$, the assumption that $g$ is analytic does introduce
a link between this behavior and the behavior at $E<E_{c}$ . In particular,
if we assume a finite value for $\sigma_{\infty}$, this implies that
$d^{2}g/d(E^{2})^{2}\to0$ for large $E^{2}$, while it is positive
at $E=E_{c}$, which at least suggests that $d^{3}g/d(E^{2})^{3}$
is likely to be negative at $E=E_{c}$. If we assume that $g$ has
the form of a cubic polynomial in $E^{2}$ for $E<E_{c}$, this would
imply that $g(0)-g(E_{c})$ is larger than in case of a pure quartic
form, for the same values of $E_{c}$ and $\sigma_{c}$.

In previous work, (Refs. \onlinecite{ZRS-ours1,ZRS-ours2}), we
assumed the following form for $g$:

\begin{equation}
g_{\beta}(\be)=\sigma_{c}\frac{1+\beta^{2}}{4}E^{2}-\sigma{}_{c}E_{c}^{2}\frac{\left(1+\beta^{2}\right)^{2}}{4}\ln\left(1+\frac{E^{2}}{\left(\beta E_{c}\right)^{2}}\right).\label{eq:g from PRL}\end{equation}
 The high field conductivity is then given by $\sigma_{\infty}=\sigma_{c}(1+\beta^{2})/2$.
In the limit $\beta\rightarrow\infty$, the function $g_{\beta}$
reduces to the quartic form $g_{quartic}(\be)$ given by (\ref{eq:gquad}).
We note that $\beta=0.1$ was used in numerical calculations whose
results are shown in Figs. 1 and 2 of Ref. \onlinecite{ZRS-ours1}
and Figs. 4 and 5 of Ref. \onlinecite{ZRS-ours2}

We shall now examine the way in which several different assumptions
for $g$ affect the relative Lypunov cost of domain walls of different
angles $\gamma$. We first demonstrate that, for the quartic form
used in our calculations, the Lyapunov energy of a domain wall is
proportional to $\sin^{3}\gamma$. We will consider an infinite system,
with a single domain wall centered around $y=0$. The problem is invariant
under translations along x-axis, so that electric field $\be$, and,
consequently, longitudinal current $\bj^{l}$, are functions of y-coordinate
only. Moreover, the x-component of field is constant: $E_{x}=E_{c}\cos\gamma$.
The boundary conditions on the y-component are $E_{y}(\pm\infty)=\pm E_{c}\sin\gamma$.
We are looking for a time-independent solution of a continuity equation.
Since both electric field $\be$ and Hall conductivity $\sigma_{H}$
depend on $y$ only, the Hall current has a vanishing divergence.
Therefore, solving $\mathrm{div}\,\bj=0$ would only entail solving
$dj_{y}^{l}(y)/dy=0$, where the longitudinal current $\bj^{l}$ is
given in Eq. (\ref{eq:long current}).

Since $|\be|=E_{c}$ at $y=\pm\infty$, the current density vanishes
far away from the domain wall. It follows that $j_{y}^{l}$ must vanish
for all $y$. Then, the equation to solve is \begin{eqnarray}
j_{y}^{l} & = & \sigma_{c}\frac{E^{2}-E_{c}^{2}}{2E_{c}^{2}}E{}_{y}-\lambda\nabla^{2}E_{y}\nonumber \\
 & = & \sigma_{c}\frac{E_{y}^{2}-E_{c}^{2}\sin^{2}\gamma}{2E_{c}^{2}}E_{y}-\lambda\nabla^{2}E_{y}=0.\label{eq:1D eqn}\end{eqnarray}
This is one of the few nonlinear differential equations that happens
to be exactly solvable. The solution that satisfies both the boundary
conditions and the requirement that the domain wall be centered around
$y=0$ is \begin{equation}
E_{y}(y)=E_{c}\sin\gamma\tanh\left(\frac{\sin\gamma}{\sqrt{\lambda/\sigma_{c}}}y\right)\label{eq:Quartic Field}\end{equation}
Using this field profile, we find the Lyapunov cost per unit length
of domain wall is given by \begin{eqnarray}
\delta G(\gamma) & = & \int_{-\infty}^{\infty}[g(\be)+\frac{\lambda}{2}(\nabla\cdot\be)^{2}-g(E_{c})]dy\nonumber \\
 & = & \frac{2\sqrt{\lambda\sigma_{c}}E_{c}^{2}\sin^{3}\gamma}{3}.\label{eq:Gdw}\end{eqnarray}

\begin{figure}
\includegraphics[width=70mm,keepaspectratio]{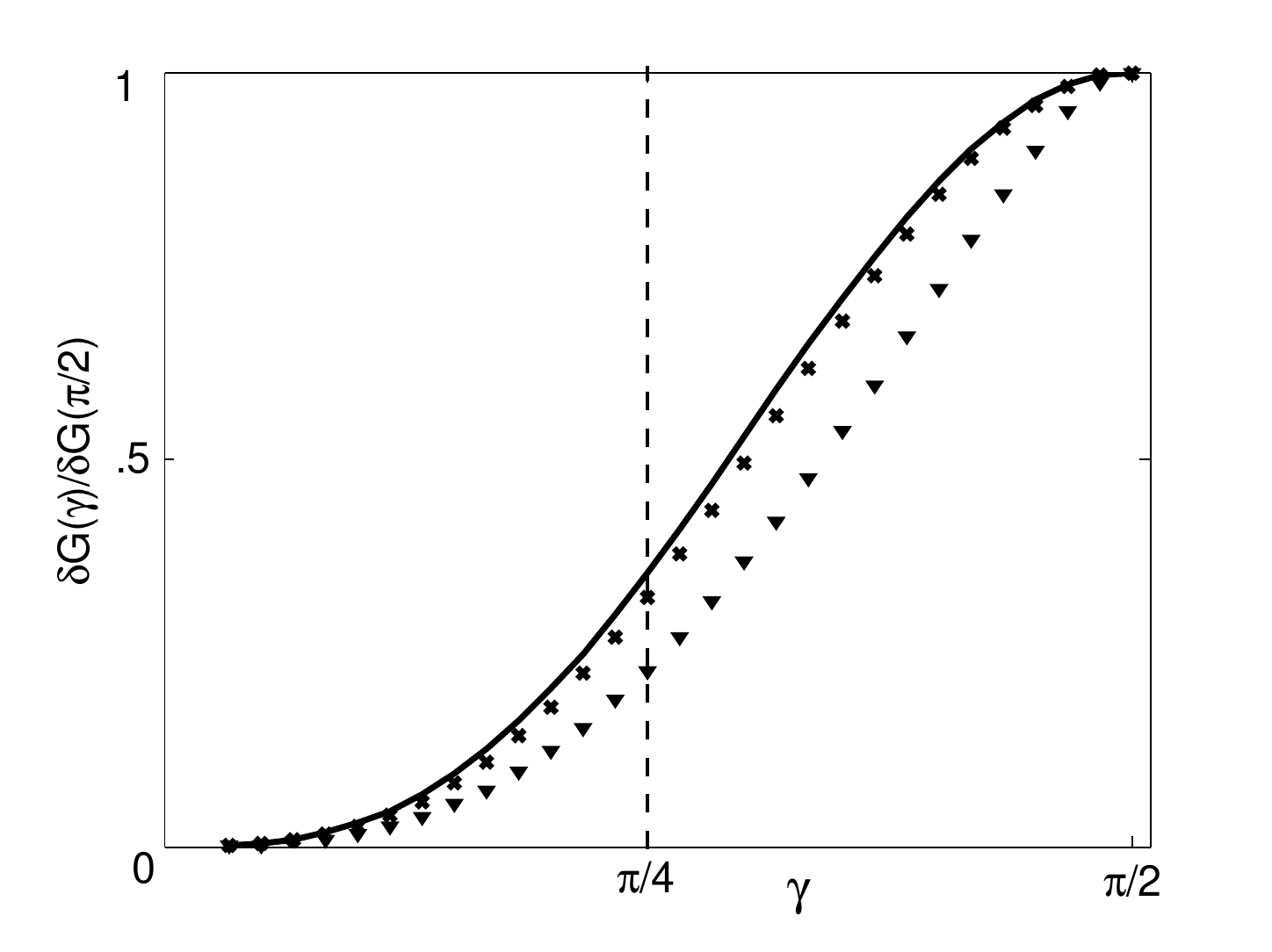}

\caption{\label{fig:domainenergy} Normalized Lyapunov cost of domain wall
per unit length, ${\delta G(\gamma)}/{\delta G(\pi/2)}$, as a function
of the angle $\gamma$ between the field in the bulk of the domain
and the domain wall. Solid line is the exact result for quartic Lyapunov
density. Triangles correspond to the Lyapunov density $g_{\beta}$
of Eq. \ref{eq:g from PRL} with the value $\beta=0.1$, used in our
previous work (Refs. \onlinecite{ZRS-ours1,ZRS-ours2}), while crosses
show results for $\beta=1.$}
\end{figure}

A closed form expression cannot be obtained for the Lyapunov density
given by Eq. \ref{eq:g from PRL}. In that case, we can solve $j_{y}^{l}(y)=0$
numerically, then integrate to get $\delta G(\gamma)$. Figure \ref{fig:domainenergy}
shows the Lyapunov cost per unit length of a domain wall as a function
of angle $\gamma$, normalized by its value at $\gamma=\pi/2$, for
quartic Lyapunov density $g_{quartic}(\be)$, as well as for $g_{\beta=.1}(\be)$
(used in Refs. \onlinecite{ZRS-ours1,ZRS-ours2}) and $g_{\beta=1}(\be),$
which gives $\sigma_{\infty}=\sigma_{c}$.

Two points need to be made. First, we note that the ratios $\delta G(\pi/4)/\delta G(\pi/2)$,
for both $\beta=0.1$ and $\beta=1$, are smaller than the value $2^{-3/2}$
that is obtained from the quartic form. Thus as stated in Sec. \ref{sec:Uniform-Hall-Conductance},
Fig. \ref{fig:domainenergy} demonstrates that more realistic forms
of Lyapunov density would break the degeneracy between having a single
horizontal domain wall and a couple of diagonal ones, with diagonal
walls (corresponding to $\pi/4$ domains) being preferred at least
in these examples. That, of course, is helpful, because it is the
time evolution of $\pi/4$ walls that produces dynamics.

The second point is that the energy cost of very small angle domain
walls is actually universal, independent of the the precise form of
$g(\be)$, if $E_{c}$ and $\sigma_{c}$ are specified. Indeed, if
the angle $\gamma\ll1$, then even the smallest value that $(E/E_{c})^{2}$
attains, $\cos^{2}\gamma$, is very close to 1. But, by construction,
all $g(\be)$ look alike in the neighborhood of $E=E_{c}$. Hence,
the energy penalty for a small angle domain wall is given by (\ref{eq:Gdw}),
irrespective of form of $g(\be)$.

%\bibliographystyle{apsrev} \bibliographystyle{apsrev}
%\bibliography{\string"C:/Physics Projects/Thesis/refs\string"}

\end{document}